\documentclass[reprint,showpacs,preprintnumbers,API,prb]{revtex4-1}
\usepackage{graphicx}
\usepackage{dcolumn}
\usepackage{bm}

\preprint{API/123-QED}

\begin{document}
	\title{Hairy BTZ black hole and its analogue model in graphene}
	\author{B. S. Kandemir}
	\email{kandemir@science.ankara.edu.tr}
	\address{Department of Physics,
		Ankara University, Faculty of Sciences, 06100, Tando\u gan-Ankara,
		Turkey\\}
	
	\date{\today}

\begin{abstract}
We obtain a novel exact analytical solution to Einstein-Maxwell-scalar gravity with negative cosmological constant in (2+1)-dimensions. The scalar field is minimally coupled to gravity and electromagnetism by the metric ansatz. On the one hand, we  first find analytical results for  the real part of  quasinormal mode spectrum of a rotating hairy   Ba\~{n}ados-Teitelboim-Zanelli (BTZ) black hole by solving the associated Dirac equation through the pseudo-Hermitian quantum mechanical tools within the framework of
discrete-basis-set method.   We then show that, in the spinless case, these quasinormal modes (QNMs)  have equally spaced with respect to the related azimuthal quantum number, while the spinning  BTZ black hole with scalar hair has Dirac-like unequally spaced discrete one. Moreover, the analytical results for these QNMs found here are  in excellent agreement with those found in the literature for hairless BTZ black holes.
On the other hand, we also develop an analogue model for this hairy Banados-Teitelboim-Zanelli black hole in graphene, by mapping it onto the hyperbolic pseudosphere surface with negative curvature. The model not only offers to make some predictions for the gravity-like phenomena in a curved graphene sheet but also paves the way for reproducing black hole thermodynamics scenarios in two-dimensional topological insulators.
\end{abstract}

\maketitle

\address{Department of Physics,
Ankara University, Faculty of Sciences, 06100, Tando\u gan-Ankara,
Turkey\\}

\section{Introduction}

After the discovery of Banados-Teitelboim-Zanelli (BTZ)\cite{Banados92}  black hole, gravity in 2+1 dimensional spacetime not only provide a new window for a deeper understanding of higher dimensional gravities but also offers different perspectives to realize table top experiments in 2D condensed matter systems\cite{Unruh,Volovik},  particularly, Hawking–Unruh phenomenon\cite{Bekenstein99,Hawking75,Dreyer03,Hod98}  in graphene\cite{Iorio11,Iorio12a,Iorio12b,Iorio13,Iorio14,Cvetic12}(See for a review Ref.~(\onlinecite{Amorima2016})).

In the context of 2+1 dimensional gravity, there is a large literature on 
quasinormal modes and 
associated frequencies for 3D spinless BTZ model\cite{Mann97,Mann99,Cardoso01,Birmingham01,Birmingham02,konoplya04, Birmingham04,Myung12,Becar14}. Despite these efforts, there is still need a self consistent theory for the quasinormal modes of  hairy BTZ black hole. It should also be remembered that all the studies made so far in the case of hairless BTZ black holes are concerned with the spinless and hairless BTZ black holes and thus achieving quasinormal mode spectrum for the hairy BTZ black hole is still a challenging problem itself.

In our recent work\cite{kandemir}, we find quasinormal modes of BTZ black hole and proposed a variational approach to the solutions for the associated Dirac Hamiltonian of graphene pseudo-particles in BTZ black hole background. The approach introduced is an attempt to address how the complicated aspects of (2+1)-dimensional gravity can be simulated in a Beltrami trumpet shaped graphene. Beltrami trumpet is a two-dimensional surface whose metric is conformally equivalent to the BTZ black hole solution of (2+1)-dimensional Einstein gravity with a negative cosmological constant, $\Lambda=-1/l^2$. In other words, the out of horizon part of the BTZ metric can be embedded into three dimensional space, and hence it is conformally related to the Beltrami trumpet surface. Since the massless Dirac equation has also conformal symmetry, the conformal equivalnece of BTZ black hole and Beltrami trumpet metrics result with Dirac particles moving on gravitational backgrounds. This means that the properties of Dirac pseudo-particles on the Beltrami trumpet shaped graphene sheet can be used to obtain a physical realization of these particles moving on the BTZ black hole background. 

In this paper, we suggest a model in which the coupling to Einstein-Maxwell scalar gravity with negative cosmological constant is taken into account. To do this we first write the associated Dirac Hamiltonian from the considerations of conformal relevances of BTZ metric in the presence of electromagnetic vector potential and Beltrami trumpet metric. Then, to cope with the non-Hermitian character of the resultant Hamiltonian we use pseudo-Hermitian quantum mechanical tools. Finally, we introduce a discrete-basis-set method to obtain its solutions to avoid from variational collapse due to the existence of both positive and negative eigenstates of Dirac Hamiltonian. Therefore, we obtain not only exact analytical results for  the real part of  quasinormal mode spectrum of a rotating hairy   BTZ black hole but also construct  its an  analogue  model in a curved graphene sheet which allows to make some predictions for the gravity-like phenomena.

\section{Theory}

The charged BTZ black hole solution of the 3D Einstein gravity with negative cosmological constant is given by the
axially symmetric metric
\begin{equation}
ds^{2}=-\Delta (r)dt^{2}+\Delta ^{-1}(r)dr^{2}+r^{2}\left( d\phi - w(r)dt\right) ^{2}  \label{1}
\end{equation}
with the lapse function $\Delta (r)$ and the angular shift $w(r)$ 
\begin{eqnarray}
\Delta (r)&=&-M\left(1+\frac{2B}{3r}\right)+\frac{r^2}{l^2}+\frac{\left(3r+2B\right)^2J^2}{36r^4} \nonumber
\end{eqnarray}
and
\begin{eqnarray}
w(r)&=& \frac{(3r+2B)J}{6r^3}. \nonumber
\end{eqnarray} 
Here, $M$, $J$ and $B$ are integration constants associated with asymptotic invariance under time displacements (mass), rotational invariance (angular momentum) and free parameter characterizing the scalar field, respectively, and $l$ is the cosmological length that is related to
the negative cosmological constant $\Lambda $ by $l=\sqrt{-\Lambda }$.

Since the static metric can be conformally rescaled up to a
conformal factor $\Delta (r)$, it is easy to show that its scaled ultra
static part is conformal to the axisymmetric optical Zermelo metric,
\begin{eqnarray}
  g_{ij}=\left[\begin{array}{ccc}
    -1+r^2w^2(r)/\Delta(r) & 0 & -r^2w(r)\\
    0& 1/\Delta^2(r) &0\\
    0& 0& r^2/\Delta(r)
\end{array}\right]\nonumber
\end{eqnarray}
with its inverse
\begin{eqnarray}
   g^{ij}=\left[\begin{array}{ccc}
       -1 & 0 & -w(r) \\
        0 & \Delta^2(r) &0 \\ 
        -w(r) & 0 & -w^{2}(r)+\Delta(r)/{r^2} 
\end{array}\right]\nonumber
\end{eqnarray}
It can be mapped onto Beltrami trumpet surface with constant
negative curvature. This conformal relation allows one to study the effects
of horizons on low energy excitations of massless Dirac fermions on a curved
graphene sheet by considering the associated massless Dirac equation
subjected to externally applied electric field and gravito-magnetic field in
the BTZ metric background. The resultant Dirac equation in the optical BTZ
black hole background can be written  as%
\begin{eqnarray}
&&\left[ \sigma _{1}\left( \Delta \frac{\partial }{\partial r}-\frac{M}{2r}+%
\frac{J^{2}}{4r^{3}}\right) +\sigma _{3}\frac{im}{r}\Delta ^{1/2}\right. 
\nonumber \\
&&\qquad \qquad \quad \left. -\sigma _{2}\left( -\epsilon+m\frac{J}{2r^{2}}%
\right) -\frac{J}{4r^{2}}\Delta ^{1/2}\right]\psi (r) =0,  \nonumber \\
&&  \label{3}
\end{eqnarray}%
where $\sigma _{i}$ are the Pauli spin matrices, and axial symmetry in
stationary picture is taken into account by choosing the wave function $\Psi
=\psi (r)e^{-i\epsilon t+im\phi }$ with $m$ is the azimutal angular quantum
number.
\begin{table}
	\caption{Components of $H_{\nu \underline{\epsilon };\alpha }$.}
	\resizebox{9.1cm}{!}{
		\centering
		\begin{ruledtabular}
			\begin{tabular}{ c  c  c c c c }
				$\underline{\epsilon }$ & $\alpha$ & $\nu$ & $0$ & $1$ & $2$ \\ [1ex] \hline
				\hline
				$0$ &   &  & $0$ & $-\frac{r^2w(r)w(r)^{\prime}}{2\Delta(r)}$ & $0$ \\ [2ex]
				$1$ &$0$&  & $0$ & $\frac{1}{2\Delta(r)\sqrt{\Delta(r)}}G(r)$ & $0$ \\ [2ex]
				$2$ &   &  & $-\frac{rw}{2\Delta}F(r)$ & $0$ & $-\frac{r}{2\Delta(r)}H(r)$ \\ [2ex]
				\hline
				$0$ &   &  &$-\frac{r^2w(r)w^{\prime}(r)}{2\Delta(r)}$ & $0$ &$-\frac{r^2w^{\prime}(r)}{2\Delta(r)}$ \\ [2ex]
				$1$ &$1$&  &$-\frac{rw^{\prime}(r)}{2\sqrt{\Delta(r)}}$ & $0$ & $0$ \\ [2ex]
				$2$ &   &  & $0$ & $0$ & $0$ \\ [2ex]
				\hline
				$0$ &   &  & $0$ & $\frac{r^2w^{\prime}(r)}{2\Delta(r)}$ & $0$ \\ [2ex]
				$1$ &$2$&  & $0$ & $-\frac{2\Delta-r\Delta(r)^{\prime}}{2\Delta(r)\sqrt{\Delta(r)}}$ & $0$ \\ [2ex]
				$2$ &   &  & $-\frac{r}{2\Delta(r)}H(r)$ & $0$ & $-\frac{r (-2\Delta(r)+r\Delta(r)^{\prime})}{2\Delta(r)}$\\ [2ex]
			\end{tabular}
		\end{ruledtabular}
	}
\end{table}\\

In general, the associated Hamiltonian describing the dynamics of spin $1/2$
particles in a gravitational background such as in Eq.~(\ref{3}) has no
hermiticity. Fortunately, it is shown very recently that, the hermiticity
problem of Hamiltonians corresponding to a spin $1/2$ particle in an axially
symmetric stationary gravitational background can be handled by using the
pseudo-Hermitian quantum mechanical tools \cite{Gorbatenko10, Gorbatenko11}.
If one finds an invertible operator $\rho $ satisfying the Parker weight
operator relationship $\rho =\eta ^{\dagger }\eta $, then it is easy to find
a Hermitian Hamiltonian such as $\mathcal{H}_{\eta }=\eta \mathcal{H}\eta
^{-1}=\mathcal{H}_{\eta }^{\dagger }$ whose spectrum coincides with those of 
$\mathcal{H}$. The Hermitian Hamiltonian can be written in terms of metric
components $g_{ab}$ together with the determinant of the metric $g$  as follows 
\begin{eqnarray}
\mathcal{H}_{\eta } &=&\mathcal{H}_{0}-\frac{i}{4(-g^{00})}\widetilde{\gamma 
}^{0}\widetilde{\gamma }^{k}\left[ \frac{\partial \ln (-g)}{\partial x^{k}}+%
\frac{\partial \ln (-g^{00})}{\partial x^{k}}\right]  \nonumber \\
&&\qquad+\frac{i}{4}\left[ \frac{\partial \ln (-g)}{\partial t}+\frac{%
\partial \ln (-g^{00})}{\partial t}\right]  \label{4}
\end{eqnarray}%
with 
\[
\mathcal{H}_{0}=\frac{i}{\left( -g^{00}\right) }\widetilde{\gamma }^{0}%
\widetilde{\gamma }^{k}\frac{\partial }{\partial x^{k}}-i\widetilde{\Phi }%
_{0}+\frac{i}{\left( -g^{00}\right) }\widetilde{\gamma }^{0}\widetilde{%
\gamma }^{k}\widetilde{\Phi }_{k} 
\]%
where 
\begin{eqnarray}
\widetilde{\gamma }^{\alpha } &=&\widetilde{H}_{\underline{\beta }}^{\alpha
}\gamma ^{\underline{\beta }}  \label{5} \\
\widetilde{\Phi }_{\alpha } &=&-\frac{1}{4}\widetilde{H}_{\mu }^{\underline{%
\epsilon }}\widetilde{H}_{\nu \underline{\epsilon };\alpha }\widetilde{S}%
^{\mu \nu }  \label{6}
\end{eqnarray}%
are curved space gamma matrices and transformed bispinor connectivity,
respectively, together with $\widetilde{S}^{\mu \nu }=\frac{1}{2}\left( \widetilde{%
\gamma }^{\mu }\widetilde{\gamma }^{\nu }-\widetilde{\gamma }^{\nu }%
\widetilde{\gamma }^{\mu }\right) $, and $x^{k}$, $t$ are the coordinates. In Eqs.(\ref{4}-\ref{6}), $\widetilde{%
\gamma }^{\alpha }$ are determined in terms of Dirac matrices with local
indices $\gamma ^{\underline{\alpha }}$ through the transformed tetrad
vectors $\widetilde{H}_{\underline{\beta }}^{\alpha }$ in Schwinger gauge,
and they are also related to the Dirac matrices with global indices $\gamma
^{\alpha }$ through Schwinger functions of the corresponding metric by $%
\gamma ^{\alpha }=H_{\underline{\beta }}^{\alpha }\gamma ^{\underline{\beta }%
}$. Here, the indices take values from $0$ to $2$, and semicolon denotes the
covariant derivative which is written as 
\begin{equation}
H_{\nu \underline{\epsilon };\alpha }=\frac{\partial H_{\nu \underline{%
\epsilon }}}{\partial x^{\alpha }}-\Gamma _{\nu \alpha }^{\lambda
}H_{\lambda \underline{\epsilon }}  \nonumber
\end{equation}%
where $H_{\lambda \underline{\epsilon }}=g_{\lambda
\mu }H_{\underline{\epsilon }}^{\mu }$. $\Gamma _{\nu \alpha}^{\lambda }$ denotes the
connection coefficients.
So, the calculated components of Eq.~(6) are presented in Table 1.
Thus, by using the components of the Zermelo metric conformally related to Eq.~(\ref{1}), one finds the components
of the transformed bispinor connectivity given by Eq.~(\ref{6}) as 
\begin{eqnarray}
\widetilde{\Phi }_{0} &=&\frac{1}{8}r^2\omega(r)\omega^{\prime }(r)\left( \gamma ^{\underline{0}}\gamma ^{\underline{2}}-\gamma ^{\underline{2}}\gamma ^{\underline{0}}\right)
\nonumber \\
&&+\left(-\frac{1}{8}r \omega^{\prime }(r)\sqrt{\Delta(r)}-\frac{1}{4}\omega(r)\sqrt{\Delta(r)}\nonumber \right. \\
&&\left. +\frac{1}{8}\frac{r \omega(r)\Delta(r)^{\prime}\sqrt{\Delta(r)}}{\Delta(r)}\right) \left(
\gamma ^{\underline{1}}\gamma ^{\underline{2}}-\gamma ^{\underline{2}}\gamma
^{\underline{1}}\right) \nonumber \\
\widetilde{\Phi }_{1} &=&\frac{r \omega^{\prime }(r)\sqrt{\Delta(r)}}{4\Delta(r) }\left( \gamma ^{\underline{0}}\gamma ^{\underline{1}}-\gamma ^{\underline{1}}\gamma ^{%
\underline{0}}\right)  \nonumber \\
\widetilde{\Phi}_{2} &=&-\frac{1}{8}r^2\omega^{\prime }(r)\left( \gamma ^{\underline{0}}\gamma ^{%
\underline{2}}-\gamma ^{\underline{2}}\gamma ^{\underline{0}}\right) 
\\
&&+\left( \frac{1}{4}\sqrt{\Delta(r) }-\frac{1}{8}\frac{r\sqrt{\Delta(r) }\Delta(r) ^{\prime }}{%
\Delta(r) }\right) \left( \gamma ^{\underline{1}}\gamma ^{\underline{2}%
}-\gamma ^{\underline{2}}\gamma ^{\underline{1}}\right) .\label{7}
\nonumber  
\end{eqnarray}%
Therefore, by replacing Eq.~(7) back into Eq.~(3),\ \ the
Hermitian Dirac Hamiltonian corresponding to the BTZ black hole background
can be found as 

\begin{eqnarray}
\mathcal{H}_{\eta }&=&i\sigma _{2}\left[ \Delta(r) \frac{\partial }{\partial r}+%
\frac{\Delta(r) ^{\prime }}{2}\right] +m\omega(r) 
\nonumber \\ 
&& -\sigma _{1}m\frac{%
\sqrt{\Delta(r) }}{r}-
\sigma _{3}\frac{r\omega^{\prime }(r)\sqrt{\Delta(r)}}{2},  \label{8}
\end{eqnarray}
which yields the energy eigenvalues of   Dirac pseudoparticles on a Beltrami
trumpet shaped graphene. In Eqs.(7) and (\ref{8}), the prime shows the
derivative of the lapse function with respect to $r$ and within the
framework of 3D relativistic field theory, the Hamiltonian was written in
units of $\hbar c$, where $c$ is the velocity of light that will be replaced
by the Fermi velocity $v_{F}$ in graphene.
It is well-known that the conventional Rayleigh-Ritz variational method can
not be directly applied to the Dirac equation, due to the absence of \ both
upper and lower bounds of the associated Dirac Hamiltonian. The Dirac
Hamiltonians are not bounded from below so that spurious roots may appear
due to variational instability. This is the second obstacle that one faces
when dealing with Dirac Hamiltonians. To overcome this difficulty, Drake and
Goldman \cite{Goldman81} have firstly proposed a discrete-basis-set method
to eliminate the spurious roots. This variational procedure is based on
defining a two component radial spinor as a trial function of the form 
\begin{equation}
|\Phi \rangle =g(r)\left( 
\begin{array}{c}
a \\ 
b%
\end{array}%
\right)   \label{9}
\end{equation}%
where $a$ and $b$ are variational parameters, and $g(r)$ is an arbitrary
continuous function satisfying the conditions $\int_{0}^{\infty }g^{2}(r)dr=1
$ and $\lim_{r\rightarrow \infty }g(r)=0$. We now introduce an \textit{ansatz%
} for the normalized basis function to get an approximation to the
ground-state of the system as 
\[
g(r)=2(\sqrt{M}l)^{-3/2}re^{-r/{\sqrt{M}l}}
\]%
which satisfies the necessary constraints to avoid the spurious roots. Thus, the associated eigenvalue equation stands for the condition $(%
\mathcal{H}_{\eta }-\epsilon )|\Phi \rangle =0$ where $\epsilon $ is the
energy relative to $\hbar c$ and it is real
\[
\epsilon=\frac{\langle\Phi|\mathcal{H}_{\eta}|\Phi\rangle}{\langle\Phi|\Phi\rangle}\nonumber
\]
and this gives
\[
\epsilon\left(a^2+b^2\right)=\frac{1}{2}mJ\mathcal{P}\left(a^2+b^2\right)+\frac{1}{4}J\mathcal{Q}\left(a^2-b^2\right)+2abm\mathcal{R}.\nonumber
\]
Using this condition, and the
variation with respect to $a$ and $b$, we obtain
\begin{eqnarray}
2\epsilon a&=&amJ\mathcal{P}+\frac{1}{2}aJ\mathcal{Q}+2bm\mathcal{R}\nonumber\\
2\epsilon b&=&bmJ\mathcal{P}-\frac{1}{2}bJ\mathcal{Q}+2am\mathcal{R}.
\end{eqnarray}
The resulting set of linear equations
can be self-consistently solved if and only if the corresponding
two-dimensional characteristic determinant is set to be equal to zero
\begin{equation}
\left|\begin{array}{cc}
2\epsilon-mJ\mathcal{P}-\frac{J}{2}\mathcal{Q} & -2m\mathcal{R} \\
-2m\mathcal{R} & 2\epsilon-mJ\mathcal{P}+\frac{J}{2}\mathcal{Q} \\
\end{array}\right|=0.
\end{equation}
Finally, we find eigenvalues of $\mathcal{H}_{\eta }$ in the form 
\begin{equation}
\epsilon =\frac{1}{2}m J\mathcal{P}\pm \frac{1}{4}\sqrt{\mathrm{J}^{2}%
\mathcal{Q}^{2}+16m^{2}\mathcal{R}^{2}}  \label{10}
\end{equation}%
where $\mathcal{P}$, $\mathcal{Q}$ and $\mathcal{R}$ are defined as integrals 
\begin{eqnarray}
\mathcal{P} &=&\int_{r_{+}}^{\infty}\omega(r)g^{2}(r)dr  \nonumber \\
\mathrm{Q} &=&\int_{r_{+}}^{\infty }r\omega^{\prime}(r)\sqrt{\Delta(r)}g^{2}(r)dr
\nonumber
\end{eqnarray}%
and 
\[
\mathcal{R}=\int_{r_{+}}^{\infty }\frac{g^{2}(r)}{r}\sqrt{\Delta (r)}dr.
\]%
Here, $r_{+}$ is the  event horizon of the hairy BTZ black hole and can be found from the zeros of lapse function $\Delta(r)$, i.e., from the hexic equation in the form of
\begin{eqnarray}
36r^6-36r^4-24\widetilde{B}r^3+9\bar{J}^2r^2+12\widetilde{B}\bar{J}^2r-4\widetilde{B}^2\bar{J}^2=0\nonumber
\end{eqnarray}
which exactly gives the well-known result
\begin{eqnarray}
\bar{r}_+=\frac{r_+}{\sqrt{M}l}=\frac{1}{\sqrt{2}}\left[1+\sqrt{1-\bar{J}^2}\right]^{1/2} \nonumber
\end{eqnarray}
in the hairless case. In Eq.(\ref{10}), upon
the choice of $M$ and $J$ values, following four special cases of the hairy BTZ
black hole can easily be found:

(i) For the vacuum state ($M=0$, $J=0$), Eq.(\ref{10}) reduces to 
\begin{equation}
\epsilon=\pm \frac{m}{l}  \label{11}
\end{equation}
which is exactly equals to that found in the hairless case. Here $+$ and $-$ signs refer to left and right QNMs, respectively.

(ii) For the static black hole ($M\neq 0$, $J=0$) Eq.(\ref{10}) gives 
\begin{equation}
\epsilon=\pm \frac{m}{l}\int_{\bar{r}_{+}}^{\infty }\bar{r}^{2}e^{-2\bar{r}}\sqrt{\bar{r}^{2}-(1+\frac{2\bar{B}}{3\bar{r}})}dr  \label{12}
\end{equation}%
where  the quantities $r$, $r_{+}$ and $B$ are made dimensionless by dividing with  $\sqrt{M}\ell$, i.e.,  $\bar{r}$, $\bar{r}_{+}$ and $\bar{B}$ respectively.

(iii) For the case $M=-1$, $J=0$ which is recognized as anti-de Sitter (AdS)
spacetime, Eq.~(\ref{11}) gives 
\begin{equation}
\epsilon=\pm \frac{m}{l}\int_{\bar{r}_{+}}^{\infty }\bar{r}^{2}e^{-2\bar{r}}\sqrt{\bar{r}^{2}+(1+\frac{2\bar{B}}{3\bar{r}})}dr.
\label{13}
\end{equation}%

(iv) For the extremal BTZ black hole ($M\neq 0$, $J=Ml$) Eq.~(\ref{10})
becomes 
\begin{equation}
\epsilon=\frac{2}{l}\left[ m\,	(\mathbf{I1})\pm \sqrt{M\,	(\mathbf{I2})^{2}+m^{2}\,(\mathbf{I3})^{2}}%
\right]  \label{14}
\end{equation}
The dimensionless integrals $\mathrm{I1}$, $\mathrm{I2}$
and $\mathrm{I3}$ are given by 
\begin{eqnarray*}
	\mathbf{I1}
	&=&\int_{\bar{r}_{+}}^{\infty }d\bar{r}e^{-2\bar{r}}(1+\frac{2\bar{B}}{3\bar{r}}),\nonumber \\
	\mathbf{I2}
	&=&\int_{\bar{r}_{+}}^{\infty }d\bar{r}e^{-2\bar{r}}(1+\frac{\bar{B}}{\bar{r}})\\
	&\times&\left[-1+\bar{r}^2+ \frac{1}{4\bar{r}^{2}}-\frac{\bar2{B}}{3\bar{r}}+\frac{4\bar{B}^{2}+6\bar{r}\bar{B}}{36\bar{r}^4}\right]^{1/2} ,\nonumber \\
	\mathbf{I3} &=&\int_{\bar{r}_{+}}^{\infty }d\bar{r%
	}\bar{r}\,e^{-2r}\\
		&\times&\left[-1+\bar{r}^2+ \frac{1}{4\bar{r}^{2}}-\frac{\bar2{B}}{3\bar{r}}+\frac{4\bar{B}^{2}+6\bar{r}\bar{B}}{36\bar{r}^4}\right]^{1/2} ,
\end{eqnarray*}%
respectively.  
Here, it should be mentioned that energy eigenvalues Eq.(\ref{11}-\ref{12})
for the spinless and hairless BTZ black hole are similar to those found for the real part
of quasinormal frequencies in Refs.~ (\onlinecite{Cardoso01,Birmingham01,Birmingham04,Myung12,Becar14}). In each of the first three cases, Eqs.(\ref{11}-\ref{13}) leads to a
sequence of different ground-state energies depending on the azimuthal
quantum number. They are equally spaced below and above the Dirac point, $%
m=0 $. In the latter case, i.e., extreme black hole case, in Eq.(\ref{14})
there appears a gap which is equal to $2\sqrt{M}\mathbf{I2}/l$ together with
Dirac-like unequally spaced discrete spectrum.

In the context of graphene, by using the graphene units, Eq.~(12) can
be rewritten in terms of external electric and magnetic fields together with
a gap term as follows 
\begin{equation}
E=em\Phi _{m}(\bar{J},\bar{B},)\pm \sqrt{\Delta ^{2}(\bar{J},\bar{B})+m^{2}(\hbar \omega
_{H})^{2}},  \label{15}
\end{equation}%
where 
\begin{eqnarray}
\Phi _{m} (\bar{J},\bar{B}) &=&\frac{4\bar{J}}{\delta }\int_{\bar{r}_{+}}^{\infty}\frac{3\bar{r}+2\bar{B}}{\bar{r}}e^{-2\bar{r}}dr( \text{V}), 
\nonumber \\
\Delta (\bar{J},\bar{B}) &=&\frac{4\bar{J}}{\delta ^{2}}\bar{\mathcal{Q}}(\bar{r}_{+},\bar{J%
}, \bar{B})\,(\text{meV}),  \label{16}
\end{eqnarray}%
and frequency $\omega _{H}=\sqrt{2}v_{F}/l_{H}$ is the cyclotron frequency
in the relativistic case such that energy has   magnetic field
dependence with 
\begin{equation}
H=\frac{4H_{0}}{\delta ^{2}}\bar{\mathcal{R}}^{2}(\bar{r}_{+},\bar{J},\bar{B})(\text{T})
\label{17}
\end{equation}%
where $H_{0}=3,26\times 10^{4}$T for graphene. In Eqs.(\ref{16}) and (\ref%
{17}) , while $\delta $ denotes the ratio of cosmological length to the
lattice parameter of graphene and $\bar{J}=J/\sqrt{M}l$ is the dimensionless
spin that changes in the interval between 0 and 1. The dimensionless integrals $\bar{\mathcal{Q}}$
and $\bar{\mathcal{R}}$ are given by 
\begin{eqnarray*}
\bar{\mathcal{Q}}(\bar{r}_+,\bar{J}, \bar{B})
&=&\int_{\bar{r}_{+}}^{\infty }d\bar{r}e^{-2\bar{r}}\frac{\bar{r}+ \bar{B}}{\bar{r}}\\
 &\times&\left[\bar{r}^2-\left(1+\frac{2}{3}\frac{ \bar{B}}{\bar{r}} \right)+\frac{1}{36}\frac{\bar{J}^2}{\bar{r}^4}\left(3\bar{r}+2 \bar{B}\right)^2 \right]^{1/2} ,\nonumber \\
 \bar{\mathcal{R}}(\bar{r}_{+},\bar{J}, \bar{B}) &=&\int_{\bar{r}_{+}}^{\infty }\bar{r}d\bar{r%
}e^{-2r}\\
&\times& \left[\bar{r}^2-\left(1+\frac{2}{3}\frac{ \bar{B}}{\bar{r}} \right)+\frac{1}{36}\frac{\bar{J}^2}{\bar{r}^4}\left(3\bar{r}+2 \bar{B}\right)^2 \right]^{1/2} ,
\end{eqnarray*}%
respectively. In writing the gap term in Eq.~(\ref{16}), as a first approximation in order to
be able to write the relevant BTZ quantities in terms of graphene
parameters even for non-zero angular momentum, we take an  \textit{%
Ansatz}\cite{Iorio15,kandemir} $\sqrt{M}=1/\delta$ where $\delta=\ell/a$ is the ratio of the cosmological length to the C-C bond length in graphene. Therefore, instead of making a Beltrami trumpet shaped graphene
to observe Hawking-Unruh radiation, we model it with position dependent
electric and magnetic fields together with a position dependent mass-like
term in a flat graphene sheet.

\begin{figure}[tp]
	\includegraphics [  height=12cm,width=8.cm]{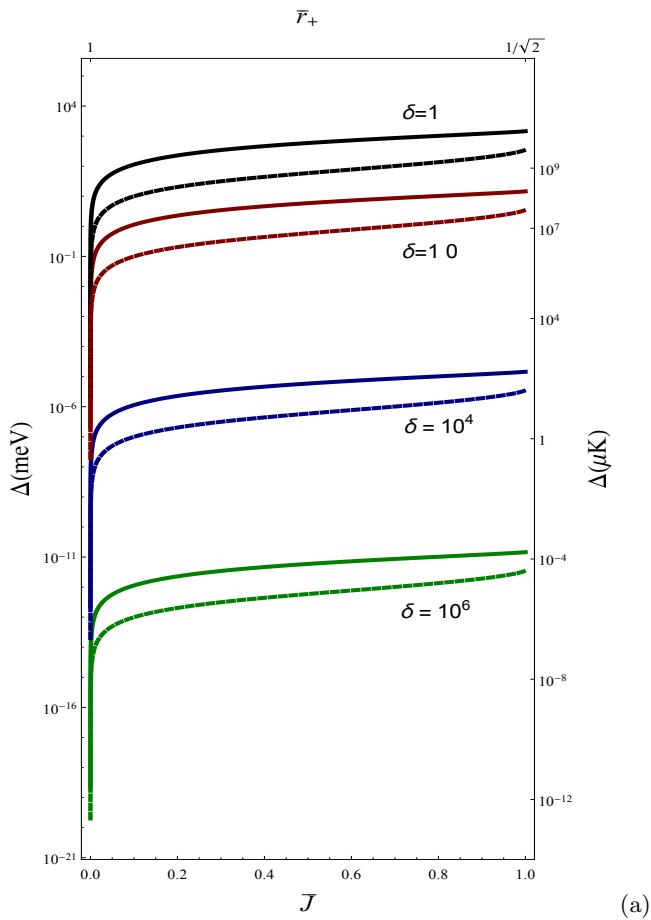} (a)
	\caption{Dependence of the half band gap as a function of the
		black hole spin for four different values of $\delta$ with $ \bar{B}=0$ (dotted lines) and with  $ \bar{B}=100/\sqrt{M}\ell$ (solid lines).}
	\label{FIG1}
\end{figure}

In FIG.(\ref{FIG1}), we plot the dimensionless spin dependence of the
resulting mass-like term, i.e., half gap for different values of $\bar{l}$
which are chosen as to be compatible with those in Ref.[\onlinecite{kandemir}], and for various of $ \bar{B}=100/\sqrt{M}\ell$. It is easy to see from the figure that the resulting gap can be
controlled via dimensionless spin. Note that the cut-off function appearing
in integrals of Eqs.(\ref{16}) and (\ref{17}) is an implicit function of
dimensionless spin $\bar{J}$ whose domain lies between 0 and 1. This implies
that $\bar{J}$ is a function of $\bar{r}_{+}$ whose range should lie between 
$1$ and $1/\sqrt{2}$.

\section{conclusions}

This paper reports for the first time a comprehensive study of  
QNMs of both spinning and spinless hairy BTZ black holes by using
pseudo-Hermitian quantum mechanical tools within the discrete-basis-set
variational method, and proposes an experimental table-top setup allowing one to
construct an analogy between gravity and condensed matter physics.
Therefore, our findings are two-fold. First, the analytical results we
obtained for the spinning hairy BTZ black hole do cover the well-known results for the hairless \cite{kandemir} and the non-rotating hairless ones \cite{Cardoso01,Birmingham02,Birmingham04,Myung12,Becar14},
and thus they are fully consistent with those found in the literature.
Here, it is important to note that our results for the real part of QNMs of hairless BTZ black hole obtained an analytical scheme proposed here are fully consistent with those found by Cardaso \textit{et al}\cite{Cardoso01} and Birmingham \textit{et al}\cite{Birmingham02} even though they used different physical approaches to calculate QNMs of hairless BTZ black hole. While Cardaso \textit{et al} calculated the associated QNMs by solving a wave equation for the scalar, electromagnetic and Weyl fields which all propogate in a spacetime with a BTZ metric, Birmingham \textit{et al} found them by conformal field theory arguments. This consistency  indicates the universal character of the real part of these QNMs and their   equally spaced spectrum, and thus these QNMs can be used to test the results found for more sophisticated situations such as hairy BTZ black hole context.

Additionally, due to the
conformal relation between the metric of the BTZ black hole and the low
energy excitations of Dirac pseudo-particles in a Beltrami shaped graphene, our model developed here allows us to propose  an experimental scheme to
observe Hawking-like effect  in a graphene sheet in the presence of uniform
magnetic and electric fields together with a gap opening term. Consequently, we showed
theoretically that different electrical and magnetic field profiles induce
different hairy black hole regimes, and they can be used experimentally to mimic  quantum properties of a 3D BTZ black hole in a laboratuary.  Finally, our conclusions could be extented to realize the same effect  in 2D topological insulators, whose bulk is insulating
while its edge is metallic, provided that, in $\delta $ which denotes the ratio of cosmological length to penetration depth length of the helical edge states, instead of the lattice parameter of the graphene. The penetration depth length  $\ell $  is expected to be of order of the lattice constant of the material which shows topological phase transition.

\begin{acknowledgements}
	This work is supported by the Scientific and Technological Research
	Council of Turkey (T\"{U}B\.{I}TAK) under the project number
	115F421.
\end{acknowledgements}

\end{document}